\RequirePackage[displaymath]{lineno}
\documentclass[twocolumn,aps,prl,showpacs,superscriptaddress,floatfix,nofootinbib]{revtex4-2}
\usepackage{newtxtext,newtxmath,booktabs,siunitx}
\usepackage{dcolumn}
\usepackage{bm}
\usepackage{float}
\usepackage{ulem}
\usepackage[]{graphicx}  
\usepackage{booktabs}
\usepackage{tabularx}
\usepackage{amsmath}
\usepackage{xcolor}
\usepackage{multirow}
\usepackage[colorlinks,citecolor=blue,urlcolor=blue,linkcolor=blue]{hyperref}

\newcommand{\four}{\{4\}}

\newcommand{\snn}{\sqrt{s_{\rm NN}}}

\newcommand{\Pb}{$^{208}$Pb}

\newcommand{\PbPb}{$^{208}$Pb+$^{208}$Pb}

\newcommand{\UU}{$^{238}$U+$^{238}$U}

\newcommand {\mean}[1]  {\langle #1\rangle}

\newcommand{\trento}{T\raisebox{-0.5ex}{R}ENTo}

\newcommand{\iebe} {{\tt iEBE-VISHNU}}

\newcommand{\puzzle}{{$v_2$-to-$v_3$ puzzle}}
\newcommand{\sig}[1]{\sigma_{\beta_{#1}}}
\newcommand{\avgoct}{\mean{\beta_{3}}}
\newcommand{\sigoct}{\sigma_{\beta_{3}}}
\newcommand{\cc}[1]{c_{#1}\{4\}/c_{#1}\{2\}^2}
\newcommand{\cce}[1]{c_{#1,\epsilon}\{4\}/c_{#1,\epsilon}\{2\}^2}
\newcommand{\vvr}{v_{2}\{2\}/v_{3}\{2\}}

\begin{document}


\title{Investigation of an octupole breathing mode of  $^{208}$Pb as a resolution of the elliptical-to-triangular azimuthal anisotropy puzzle in ultracentral relativistic heavy ion collisions}

\author{Hao-jie Xu}
\email{Corresponding author: haojiexu@zjhu.edu.cn}
\affiliation{School of Science, Huzhou University, Huzhou, Zhejiang 313000, China}
\affiliation{Strong-Coupling Physics International Research Laboratory (SPiRL), Huzhou University, Huzhou, Zhejiang 313000, China.}

\author{Duoduo Xu}
\affiliation{School of Physics, Peking University, Beijing 100871, China}

\author{Shujun Zhao}
\affiliation{School of Physics, Peking University, Beijing 100871, China}

\author{Wenbin Zhao}
\affiliation{Nuclear Science Division, Lawrence Berkeley National Laboratory, Berkeley, California 94720, USA}
\affiliation{Physics Department, University of California, Berkeley, California 94720, USA}

\author{Huichao Song}
\affiliation{School of Physics, Peking University, Beijing 100871, China}
\affiliation{Collaborative Innovation Center of Quantum Matter, Beijing 100871, China}
\affiliation{Center for High Energy Physics, Peking University, Beijing 100871, China}

\author{Fuqiang Wang}
\affiliation{Department of Physics and Astronomy, Purdue University, West Lafayette, Indiana 47907, USA}

\begin{abstract}
Relativistic heavy ion collisions provide a unique opportunity to probe the nuclear structure by taking an instantaneous snapshot of the colliding nuclei and converting it into momentum anisotropies of final emitted hadrons. A long-standing puzzle of too large a ratio of the elliptical-to-triangular (\(v_2\)-to-\(v_3\)) anisotropies in ultracentral \PbPb\ collisions at the Large Hadron Collider (LHC) cannot be solved simply by hydrodynamic simulations with initial conditions containing the spherical or certain deformed shape of \Pb.   
In this Letter, using the \iebe\ relativistic viscous hydrodynamic hybrid model simulations with the \trento\ initial condition, we show that a dynamic octupole deformation--a shape-breathing mode of  $^{208}$Pb --could potentially solve the \(v_2\)-to-\(v_3\) puzzle and simultaneously describe the  $v_3\four$ data measured in experiment. 
Our results highlight the unique capability of capturing transient collective properties of nuclei on yoctosecond (\(10^{-24}\)~s) timescales, unfeasible with low-energy nuclear reactions. 
\end{abstract}

\maketitle

{\em {\it Introduction.}} 
Relativistic heavy ion collisions at the Relativistic Heavy Ion Collider (RHIC) and the Large Hadron Collider (LHC) aim to create and study the quark-gluon plasma (QGP), a deconfined state of strongly interacting matter that once existed in the very early universe shortly after the Big Bang 
~\cite{Gyulassy:2004zy,Adcox:2004mh,PHOBOS:2004zne,BRAHMS:2004adc,Adams:2005dq,
Muller:2006ee,Jacak:2012dx,Braun-Munzinger:2015hba,Shuryak:2014zxa,ALICE:2022wpn}. 
Within the hydrodynamic framework, the pressure gradient of the QGP converts the spatial anisotropy of the initial overlapping geometry of the colliding nuclei into momentum anisotropy of final-state hadrons, called collective flow~\cite{Ollitrault:1992bk,Kolb:2003dz}. The flow anisotropies are expressed by Fourier harmonics in the azimuthal ($\phi$) distribution of final-state particles relative to the reaction planes ($\psi_n$)~\cite{Voloshin:1994mz}, 
\begin{equation}
    \frac{dN}{d\phi} \propto 1 + \sum_{n=1} 2v_n \cos n(\phi-\psi_n)\,.
\end{equation}
The success of relativistic hydrodynamics in modeling the evolution of the QGP establishes a robust connection between the initial geometry eccentricities and the final-state momentum anisotropies with $v_n \propto \epsilon_n$ for (n=2,3)~\cite{Heinz:2013th, Gale:2013da,Luzum:2013yya,Song:2013gia,Jeon:2015dfa,Song:2017wtw,DerradideSouza:2015kpt,Noronha-Hostler:2015dbi,Noronha:2024dtq}.

For ultracentral collisions of spherical nuclei at relativistic energies, the initial geometry anisotropies are predominantly driven by fluctuations, yielding comparable values, for example, for the second- and third-order eccentricities $\epsilon_{2}\approx\epsilon_{3}$~\cite{Shen:2015qta}. Because the viscous damping effects are stronger for higher-order cumulants, hydrodynamic models predict a smaller triangular flow than the elliptic flow, $v_{3} < v_{2}$~\cite{Shen:2015qta}. 
The predicted $v_{2}/v_{3}$ ratio is then too large compared to the measurements in ultracentral \PbPb\ collisions at $\sqrt{s_{\mathrm{NN}}} = 2.76\ \mathrm{TeV}$ and $5.02\ \mathrm{TeV}$~\cite{ALICE:2011ab,ATLAS:2012at,CMS:2013wjq,ALICE:2018rtz,ATLAS:2019peb}. 
This discrepancy, known as the ``ultracentral $v_{2}$-to-$v_{3}$ puzzle,'' has been a long-standing challenge in heavy ion physics for over a decade.

Nuclear structure has become a vital aspect in relativistic heavy-ion collisions, for example, in elucidating background contributions to the chiral magnetic effect in isobar collisions~\cite{Xu:2017zcn,Li:2019kkh,Xu:2021uar} and in resolving the ultracentral $v_2$ issue in  \UU\ collisions~\cite{Ryssens:2023fkv,Xu:2024bdh}.
An obvious possibility for the  anomalously large $v_3/v_2$ ratio is that the doubly magic $^{208}$Pb nucleus is octupole-deformed~\cite{Carzon:2020xwp,Zakharov:2020irp,Zakharov:2021lux}, 
with a finite $\beta_3$ value in the Woods-Saxon (WS) parameterization of nuclear density,
\begin{subequations}\label{eq:ws}
\begin{align}  
    \rho(r) &= \frac{\rho_0}{1 + \exp\left(\frac{r - R}{a}\right)}\,, \\
    R &= R_0 \left(1 + \beta_2 Y_{20} + \beta_3 Y_{30}\right)\,.  
\end{align}  
\end{subequations}\\
Here, \(R_0\) is the nuclear radius parameter, \(a\) is the diffuseness parameter, \(\beta_2\) and \(\beta_3\) quantify quadrupole and octupole deformations, respectively, and \(Y_{lm}\) denotes the spherical harmonics. 
In ultracentral collisions, the collision overlap geometry is  determined essentially by the  shape of the colliding nuclei, i.e., \(\epsilon_n^2 \propto \beta_n^2\) (for $n=$2,3)~\cite{Heinz:2004ir,Giacalone:2021udy}. 
Since \(v_n\) is proportional to \(\epsilon_n\) on the leading order, 
$v_n^2 \propto \beta_n^2$~\cite{Niemi:2012aj,Gale:2012rq,Giacalone:2021udy}.
A nonzero \(\beta_3\) would, therefore, enhance \(v_3\) and solve the \(v_2\)-to-\(v_3\) puzzle in ultracentral Pb+Pb collisions.      
However, the nonzero $\beta_3$ value fails to describe the observed \(v_3\{4\}/v_3\{2\}\) ratio~\cite{Carzon:2020xwp}.
Here $v_{n}\{m\}$ is the n-th order flow harmonics calculated with m-particle cumulant method~\cite{Poskanzer:1998yz,Borghini:2000sa,Bilandzic:2010jr}.

In this Letter, we solve the $v_{2}$-$v_{3}$ puzzle by introducing $\beta_3$ fluctuation, which implies that the colliding \Pb\ nucleus ``breathes'' in a pear shape.
The idea is as follows. It is known that $v_n\{2\}$ and $v_n\{4\}$ are  sensitive to fluctuations but in different ways; $v_n\{2\}$ measures $\sqrt{\langle v_n^2\rangle}$ and is thus sensitive to  
$\langle\beta_n^2\rangle\equiv\langle{\beta_n}\rangle^{2}+\sigma_{\beta_n}^2$, 
whereas the fluctuation effect in $v_n\{4\}$ is negative, approximately determined by $\langle{\beta_n}\rangle^{2}-\sigma_{\beta_n}^2$. The fact that, with a finite fixed $\beta_3$ value (i.e.~$\sigma_{\beta_3}=0$), the $v_3\{2\}/v_2\{2\}$ ratio is described, but not the $v_3\{4\}/v_3\{2\}$ ratio suggests that the fixed $\beta_3$ value correctly reflects $\langle\beta_3^2\rangle$, but the finite-number fluctuations are insufficient. 
With this consideration, the path to solve the \puzzle\ is clear: to enhance \(v_3\{2\}\) nonzero \(\langle \beta_3^2 \rangle\)  should be invoked; to describe the measured $v_3\{4\}/v_3\{2\}$, the $\beta_3$ value should be allowed to fluctuate to introduce extra fluctuations beyond the stochastic nucleon positions.

Such breathing modes are consistent with the findings in low-energy nuclear structure studies~\cite{Butler:1996zz,Robledo:2016xke,Poves:2019byh,Yao:2016xkd,Henderson:2025scq}.
\Pb\ stands as a benchmark nucleus in nuclear physics, playing a pivotal role in advancing our understanding of symmetry energy, a critical component of the nuclear equation of state~\cite{Li:2008gp,Horowitz:2000xj,Tsang:2012se,PREX:2021umo,Giacalone:2023cet}. 
As a doubly magic nucleus, with closed proton and neutron shells, \Pb\ exhibits exceptional stability, enabling clear observation of collective excitations like the giant monopole resonance (GMR), often termed the "breathing mode", where the nucleus uniformly expands and contracts~\cite{Ritman:1993zz,Heyde:2011pgw}. 
Beyond this fundamental mode, Pb also displays higher-order vibrational modes, such as quadrupole (L=2) and octupole (L=3) resonances, which manifest as more complex, shape-driven oscillations~\cite{Butler:1996zz,Robledo:2016xke,Yao:2016xkd,Poves:2019byh,Henderson:2025scq}. 
These modes, interpreted as higher-order "breathing" behaviors, strongly suggest that the intrinsic shapes of doubly magic nuclei such as \Pb\ must be described dynamically with the mixing of different shape configurations~\cite{Poves:2019byh,Henderson:2025scq}.

{\em {\it Model and setups.}}
In previous studies, the nuclear shape parameters, i.e. $\beta_2$ and $\beta_3$, are often treated as {\it fixed} values, and the position of nucleons are then sampled according to Eqs.~(\ref{eq:ws}). To investigate the octupole breathing mode, the  $\beta_3$ fluctuations in \Pb\ are modeled by a Gaussian probability distribution in $\beta_3$~\cite{Butler:1996zz,Henderson:2025scq},
\begin{equation}
    P(\beta_3) \propto \exp\left[-\frac{(\beta_3-\langle{\beta_3}\rangle)^2}{2\sigma^2_{\beta_3}}\right] \,.
\end{equation}
Note that there is a similar distribution along the negative direction from the symmetry-conserving mixing method shown in Ref.~\cite{Henderson:2025scq}, while the sign of \(\beta_3\) is degenerate because $v_3\{2\}$ is sensitive only to the variance of the octupole deformation $\langle\beta_3^2\rangle$.

For the following calculations, we use the \iebe\ hybrid model~\cite{Song:2010aq,Shen:2014vra} to simulate relativistic Pb+Pb collisions at $\snn=5.02$ TeV. The model contains three main components: initial conditions generated via the \trento\ model~\cite{Moreland:2014oya,Bernhard:2016tnd}; the QGP evolution modeled with relativistic viscous hydrodynamics of (2 + 1) dimension (\textsc{VISH2+1})~\cite{Heinz:2005bw, Song:2007ux, Song:2007fn}; and late stage hadronic interactions simulated with the UrQMD hadron cascade model~\cite{Bass:1998ca,Bleicher:1999xi}.   
All model parameters are taken from Ref.~\cite{Moreland:2018gsh}, calibrated for non-central Pb+Pb collisions at $\snn = 5.02$~ TeV~\cite{ALICE:2015juo,ALICE:2016ccg} without considering the nuclear deformation and neutron skin of $^{208}$Pb. 
Our investigation of the flow puzzle in the most central Pb+Pb collisions reveals that the parameters require revision.
Ultimately, to solve the flow puzzle in most central collisions, a Bayesian calibration is required to extract shape deformation and fluctuation parameters as well as transport properties such as the specific shear viscosity from the data of soft physics in Pb+Pb collisions. We leave such a work to the future; Here, we aim at a qualitative description of the data with the breathing mode of the colliding nuclei.

\begin{figure}
    \includegraphics[width=0.42\textwidth]{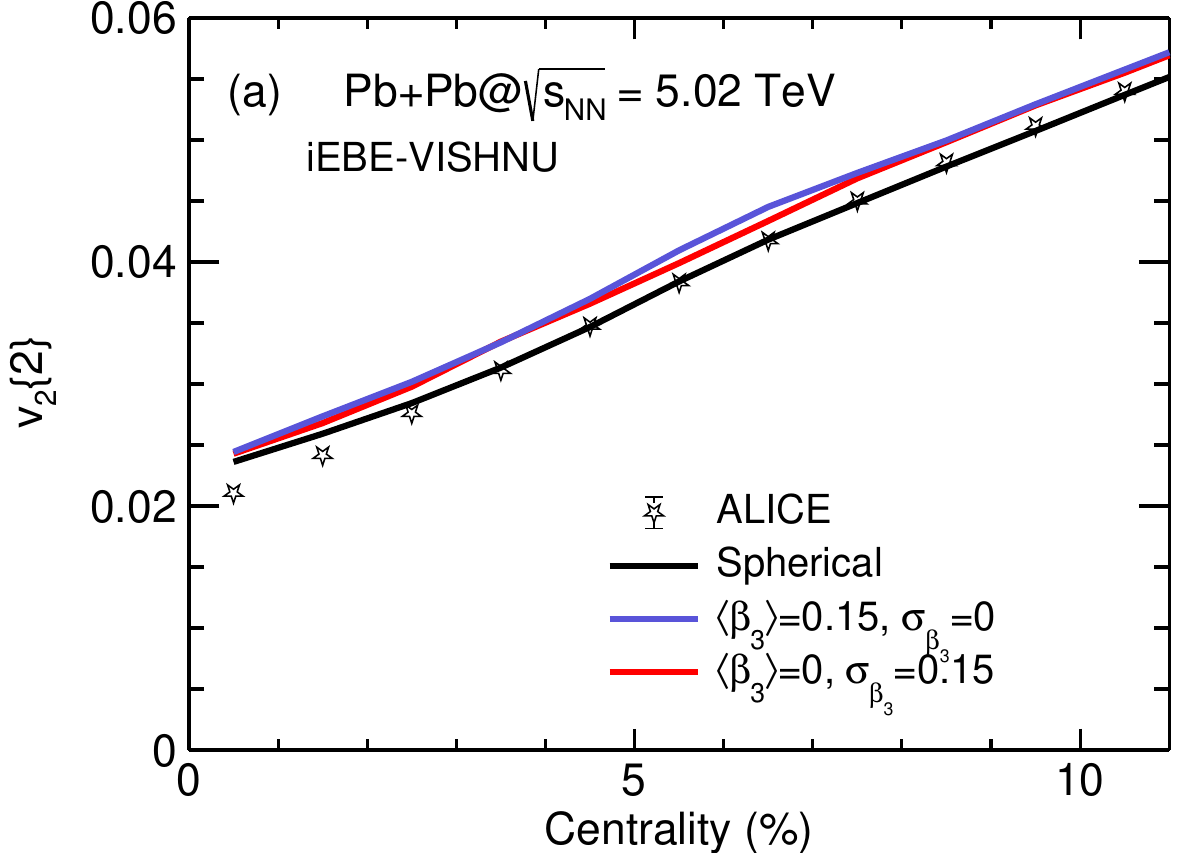}
    \includegraphics[width=0.42\textwidth]{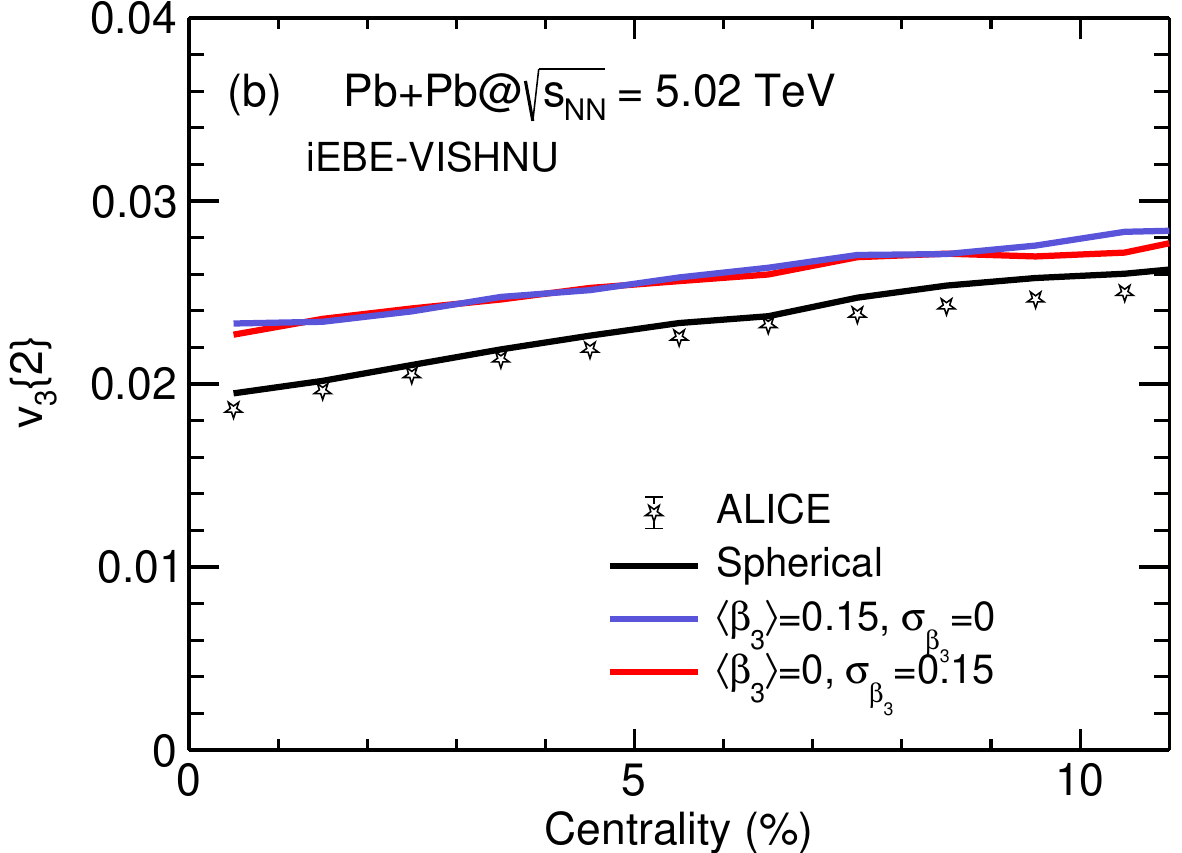}
    \caption{(Color online) 
	(a) The $v_{2}\{2\}$ and $v_{3}\{2\}$ at midrapidity in the most central Pb+Pb collisions at $\snn=5.02$ TeV, calculated from \iebe\ hybrid model with three initial Woods-Saxon geometries, together with a comparison to the data from ALICE with $|\eta|<0.8$~\cite{ALICE:2018rtz}. 
  }
\label{fig:v2v3}
\end{figure}

\begin{figure}
\includegraphics[width=0.45\textwidth]{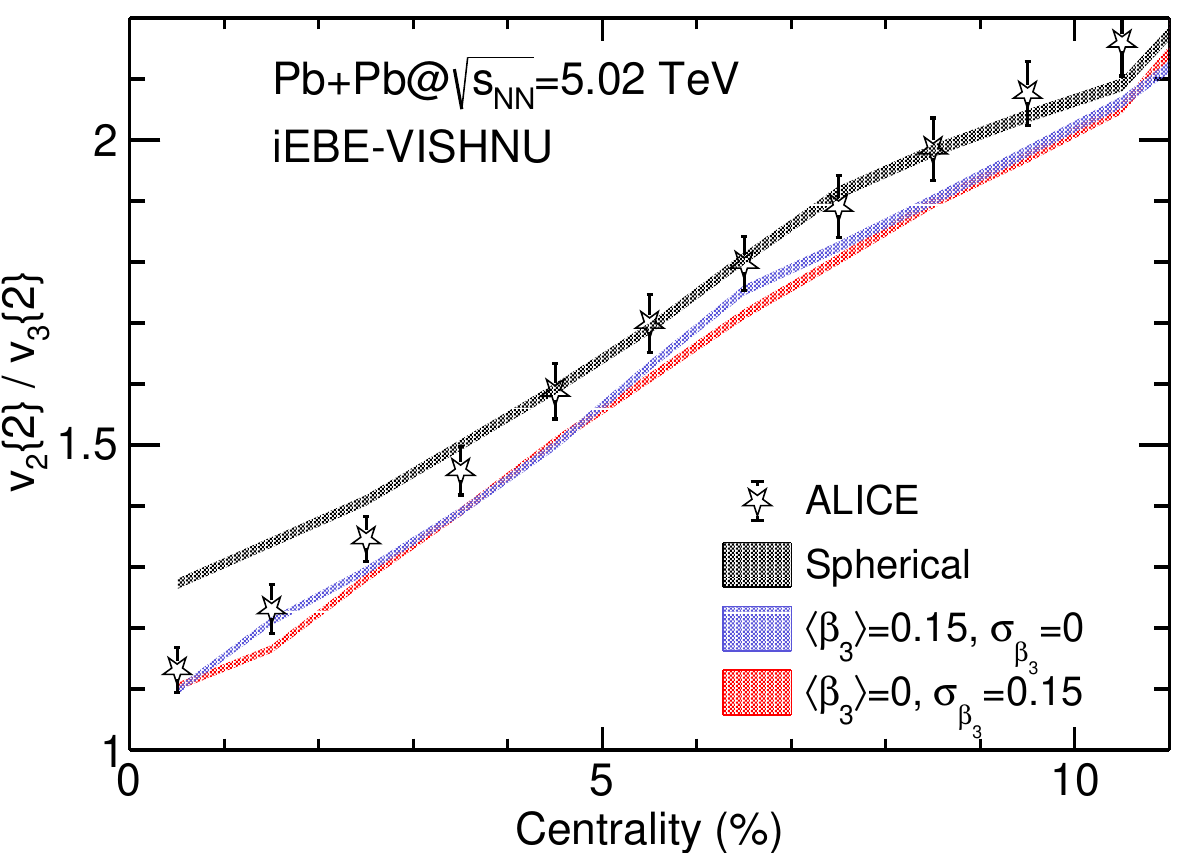}
    \caption{(Color online) 
	Similar to Fig~\ref{fig:v2v3} but for the $\vvr$ ratios at midrapidity in the most central Pb+Pb collisions at $\snn=5.02$ TeV.  The widths of the bands indicate statistical uncertainties of the calculations.}
\label{fig:flowpuzzle}
\end{figure}

{\em {\it Results and discussions.}}
In the following calculations, we firstly focus on three cases of the colliding nuclei: (i) spherical case without any deformation, (ii) a static/rigid pear-shaped \Pb\ with \(\avgoct = 0.15\) and \(\sig{3}=0\) and (iii) a "breathing" \Pb, spherical-shaped on average fluctuating into oppositely oriented pear shapes with \(\avgoct=0\) and \(\sig{3} = 0.15\). 
Figure~\ref{fig:v2v3} shows predictions for individual $v_{2}$ and $v_{3}$ in central Pb+Pb collisions at $\sqrt{s_{\text{NN}}} = 5.02~\text{TeV}$. The calibrated Bayesian parameters for a spherical \Pb, case (i), can accurately describe both $v_{2}\{2\}$ and $v_{3}\{2\}$, except for the top $2\%$ centrality. Resolving the $v_2$-to-$v_3$ ratio will require different relative changes to $v_{2}$ and $v_{3}$. This is achieved by case (ii) and (iii), though the individual $v_{2}$ and $v_{3}$ magnitudes are overpredicted.
The $v_2$ and $v_3$ magnitudes can be simultaneously tuned---for example, by adjusting bulk properties, nucleon width, or the minimum separation distance between nucleons in the colliding nuclei~\cite{Wang:2025cfu}.

Figure~\ref{fig:flowpuzzle} shows the $\vvr$ ratio in the most central Pb+Pb collisions, calculated from \iebe. Although the individual $v_{2}$ and $v_{3}$ was overestimated, both cases (ii) and (iii)  generate an enhanced magnitude of $v_3\{2\}$, which could solve the $v_2$-to-$v_3$ ratio puzzle  in ultracentral 
collisions. This result is expected because the two-particle cumulant $v_3\{2\}$ is determined only by $\langle\beta_3^2\rangle=\avgoct^2+\sigoct^2$, which is equivalent in these two cases. 
\begin{figure}
    \includegraphics[width=0.42\textwidth]{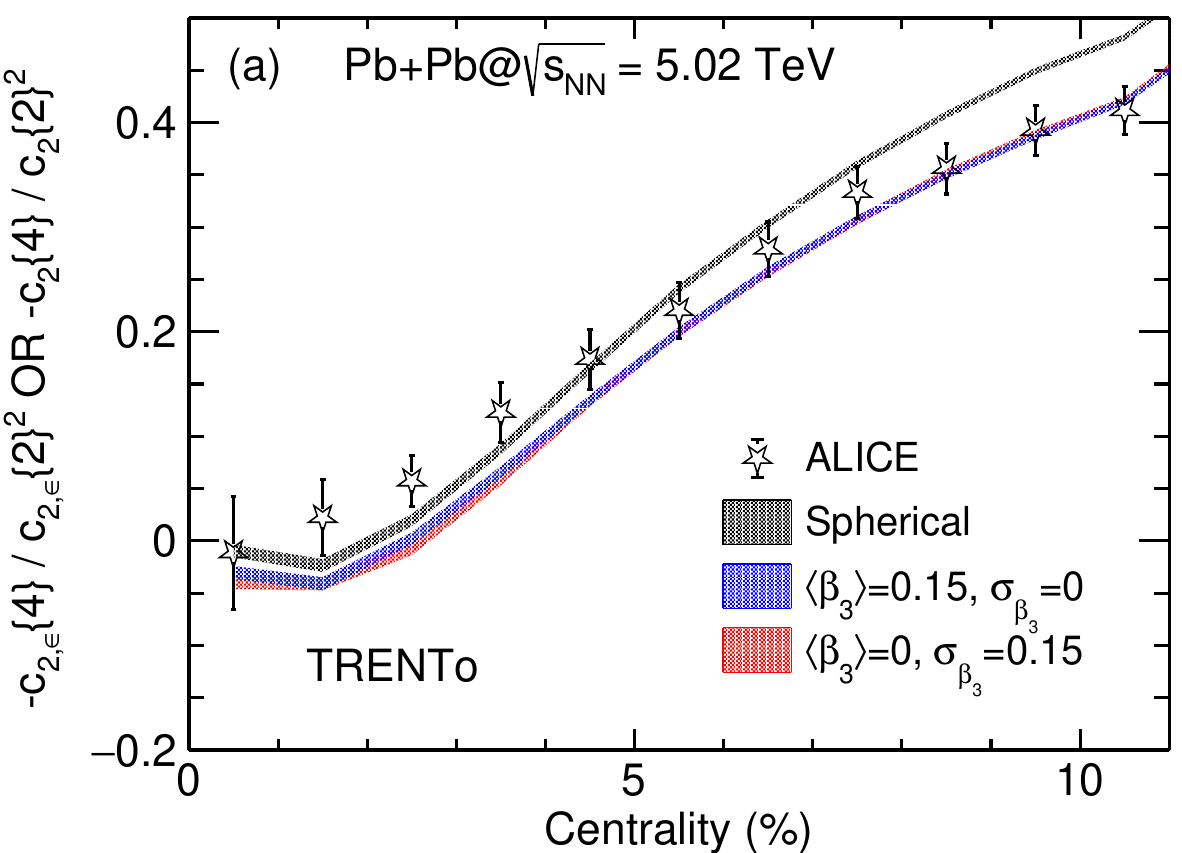}
    \includegraphics[width=0.42\textwidth]{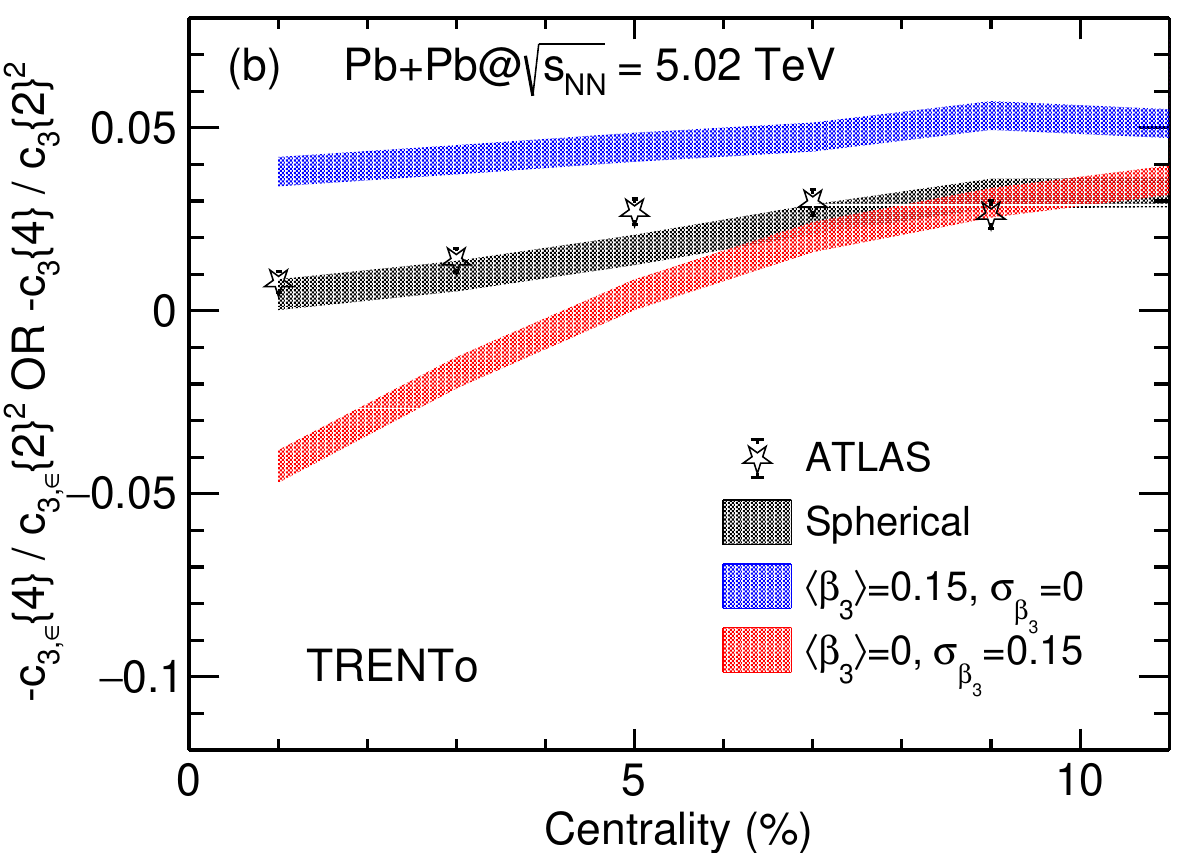}
    \caption{(Color online) Four-particle cumulants in the most central Pb+Pb collisions at $\snn=5.02$ TeV, calculated from initial-state model \trento\ with three initial geometries as used in Fig.~\ref{fig:flowpuzzle}. (a) $-\cce{2}$, compared to $-\cc{2}$ from ALICE~\cite{ALICE:2018rtz} with $|\eta|<0.8$, and (b) $-\cce{3}$, compared to $-\cc{3}$ from ATLAS~\cite{ATLAS:2019peb} with $|\eta|<2.5$.}
\label{fig:trento}
\end{figure}
  
Such an ambiguity in the two-particle cumulants can be resolved by the four-particle cumulants as briefly discussed above. However, the associated full hydrodynamic calculations are computationally expensive.
Considering the linear relationship between the final-state flow anisotropy and the initial-state eccentricity, $v_n \propto \epsilon_{n}$ for $n=2,3$ in the most central collisions~\cite{Niemi:2012aj,Gale:2012rq}, we calculate the eccentricity cumulants as a function of centrality with the \trento\ initial-state model.
Figure~\ref{fig:trento} shows the calculated ratios, $-\cce{2}$ and $-\cce{3}$, together with a comparison to the ALICE measurements $-\cc{2}$ and the ATLAS measurements $-\cc{3}$. 
Overall, the results from spherical nucleus are consistent with data (we have also verified that the residual discrepancy in $-\cce{2}$ disappears once final-state effects from hydrodynamics are considered).
This is an accident because the spherical nucleus case cannot describe the anomalously large $v_3/v_2$ ratio.
The data do not require a quadrupole (prolate or oblate) deformation of a finite $\beta_2$.  As expected, octupole deformation, with or without breathing, also does not affect the quadrupole cumulants.
On the other hand, octupole deformation directly affects \(\epsilon_3\) (and hence \(v_3\)), as shown in Fig.~\ref{fig:trento}(b). 
Neither the static deformed nor the breathing nucleus can describe the $-\cce{3}$ data.
A striking consequence of the ``breathing" mode of \(^{208}\text{Pb}\) is to drive \(-c_3\{4\}\) to significantly negative values in ultracentral. 

\begin{figure}
    \includegraphics[width=0.42\textwidth]{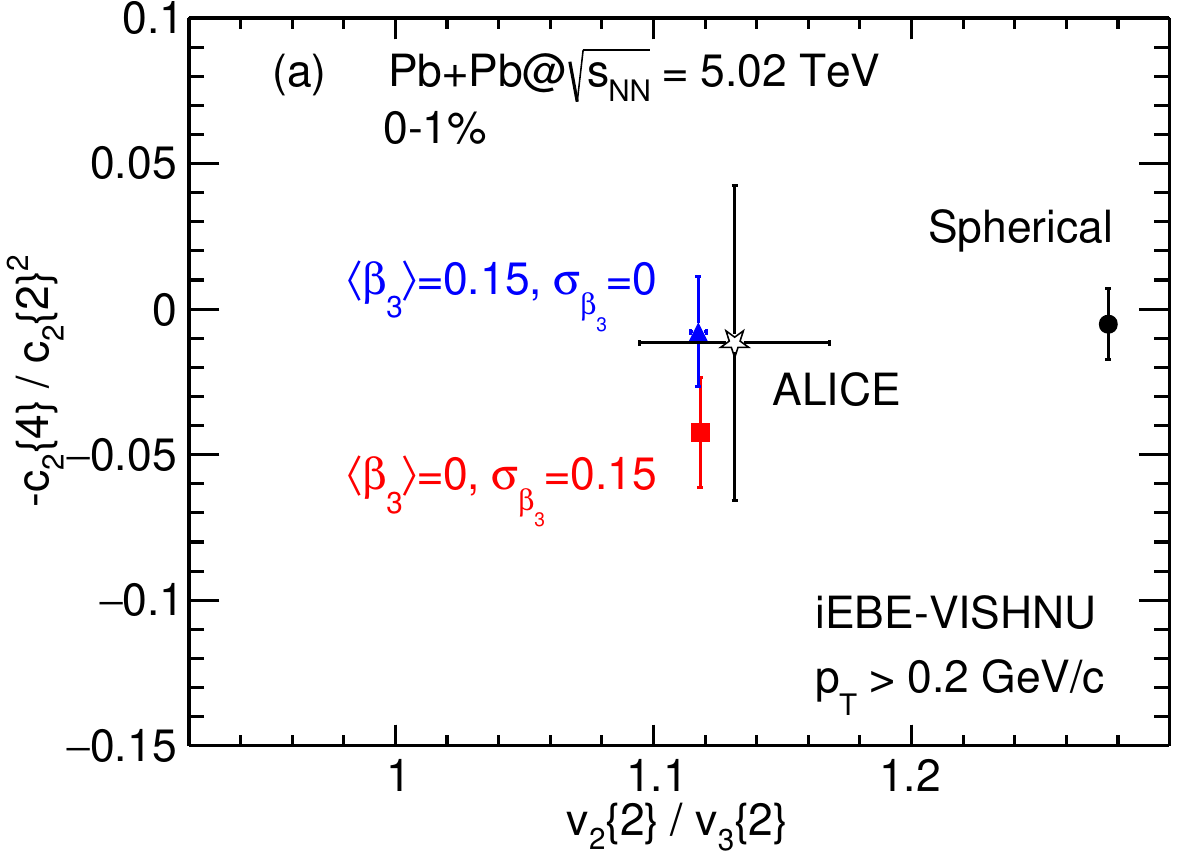}
    \includegraphics[width=0.42\textwidth]{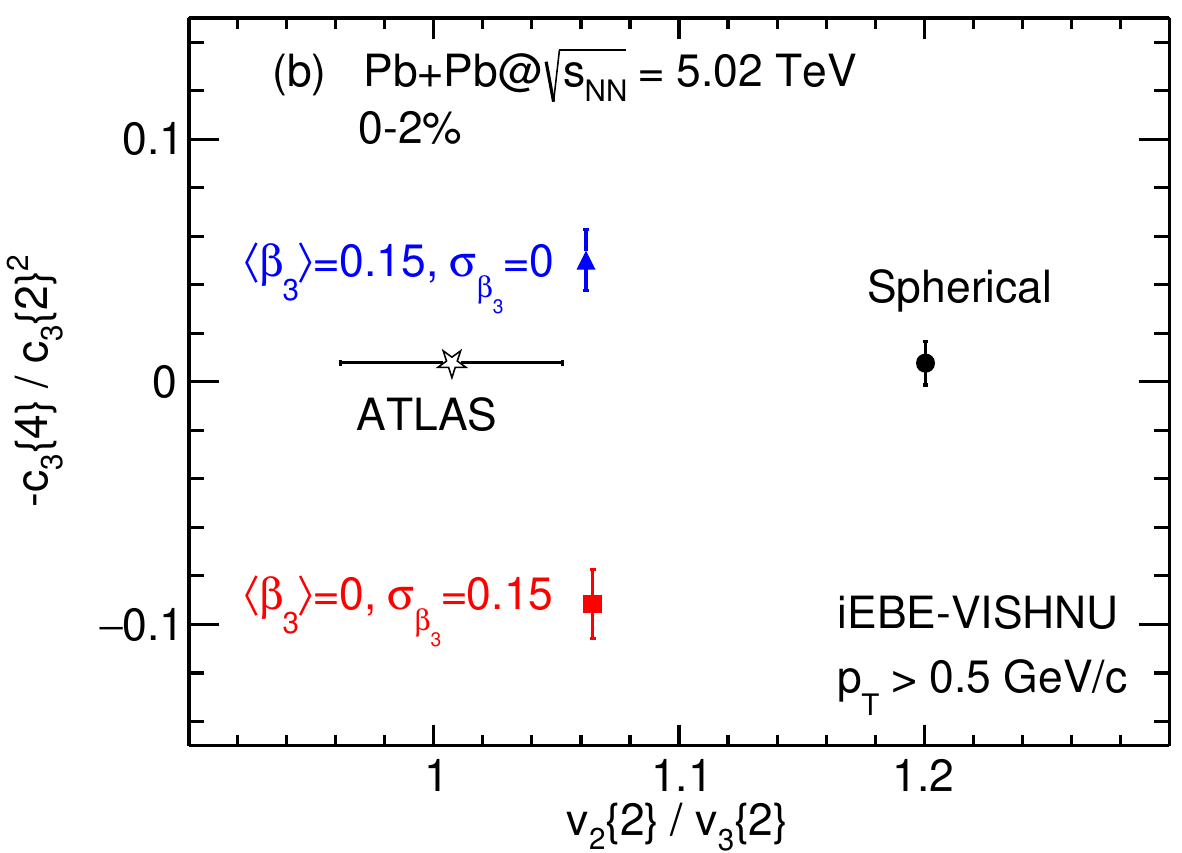}
    \caption{(Color online) 
    (a) $-\cc{2}$ vs.~$\vvr$ in the top $1\%$ centrality Pb+Pb collisions at $\snn=5.02$ TeV, calculated from \iebe\ model with three initial geometries are used as in Fig.~\ref{fig:flowpuzzle}, together with a comparison to the ALICE data~\cite{ALICE:2018rtz}, and (b) $-\cc{3}$ vs.~$\vvr$ in the top $2\%$ centrality, together with a comparison to the ATLAS data~\cite{ATLAS:2019peb}.
  }
\label{fig:vishnu}
\end{figure}

In order to identify the degree of octupole breathing of the \Pb\ nucleus, we compare final-state hydrodynamic calculations to data measurements.
The four-particle cumulants in Pb+Pb collisions have been published by ALICE for $c_{2}\{4\}$ in the top 1\% centrality~\cite{ALICE:2018rtz} and by ATLAS for $c_{3}\{4\}$ in the top 2\% centrality~\cite{ATLAS:2019peb}.
To enhance ultracentral 0--2\% collisions, we run full hydrodynamic simulations by \iebe\ with restricted impact parameter range of \(b < 4\)~fm, and then use the charged-hadron multiplicity to determine our centrality percentage (relative to minimum bias sample).
Figure~\ref{fig:vishnu} shows in (a) \(v_2\{4\}/v_2\{2\}\) within $p_{\rm T}>0.2$ GeV/$c$ in the top 0--1\% centrality compared to ALICE data, and in (b) \(v_3\{4\}/v_3\{2\}\) within $p_{\rm T}>0.5$ GeV/$c$ in the top 0--2\% centrality compared to ATLAS data.
We note that the parameter set used in this work is calibrated using anisotropic flow data as functions of centrality from ALICE~\cite{Moreland:2018gsh,ALICE:2015juo,ALICE:2016ccg}; The calculated $\vvr$ values are slightly overestimated compared to ATLAS data. In addition, the (2 + 1) dimension hydrodynamic model employed in this study may lack certain characteristics, such as longitudinal flow decorrelation over a wide pseudo-rapidity range, that may be important for the ATLAS data.

The \(v_2/v_3\) ratio is equally reproduced by the two scenarios, static deformation and shape fluctuations. 
The spherical nucleus case ($\beta_3=0$) cannot describe the data, consistent with previous conclusions~\cite{Carzon:2020xwp}. 
Although statistical uncertainties are large for the 0--1\% centrality, the \(c_2\{4\}/c_2\{2\}^{2}\) appears insensitive to $\sig{3}$, as expected.
On the contrary, \(c_3\{4\}/c_3\{2\}^{2}\) shows a strong sensitivity to $\beta_3$ deformation and fluctuations, mirroring trends in the initial-state \trento\ simulations (Fig.~\ref{fig:trento}).
Quantitatively, the final-state ratio exhibits an enhanced sensitivity, suggesting a breakdown of the simplistic mapping from initial- to final-state cumulants when shape fluctuations are involved.

\begin{figure}
\includegraphics[width=0.45\textwidth]{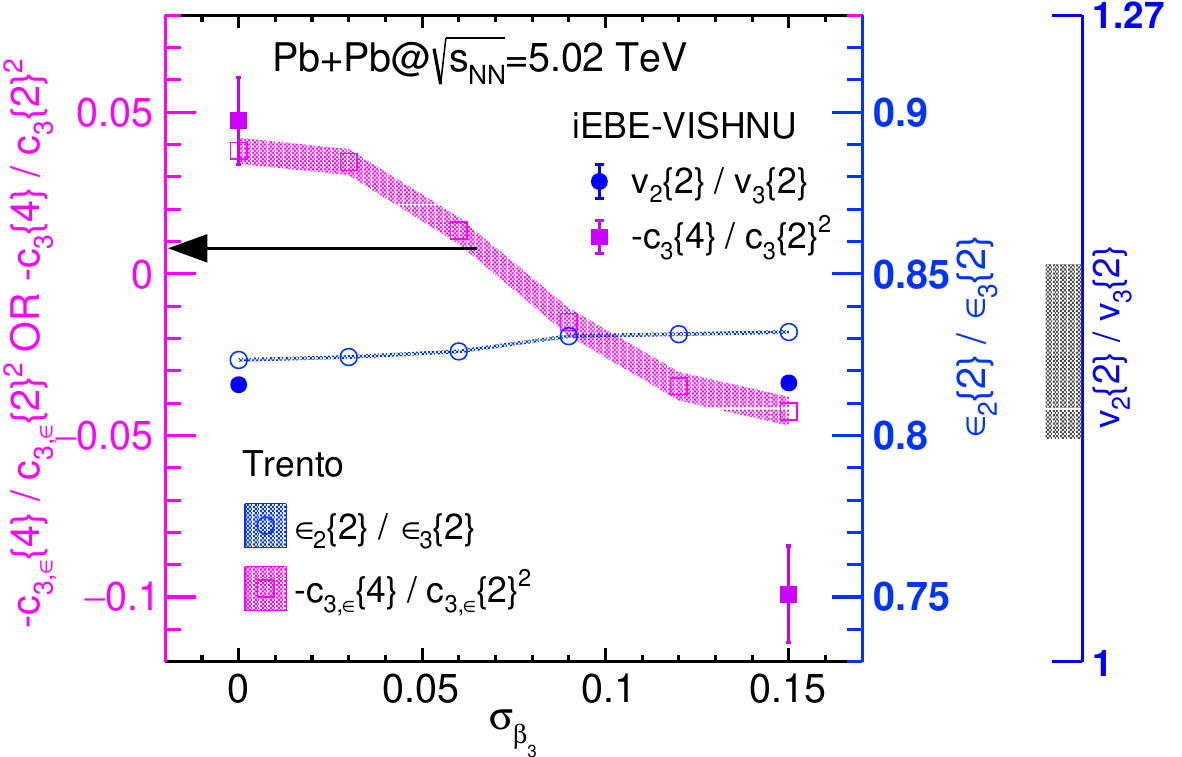}
	\caption{(Color online) $\cce{3}$ and $\cce{2}$ in ultracentral 0--1\% Pb+Pb collisions at $\snn=5.02$ TeV calculated by the initial-state \trento\ model (open points) for various combinations of  $\avgoct$ and $\sig{3}$ satisfying $\sqrt{\avgoct^{2}+\sig{3}^{2}}=0.15$. The solid points are from the final-state \iebe\ simulations for the two limiting cases with $\sig{3}^{2}=0$ and 0.15. The blue band on the extra right axis denotes the $\vvr$ data from ALICE (0--1\% centrality)~\cite{ALICE:2018rtz}, and the red arrow on the left axis denotes the $-\cc{3}$ data from ATLAS (0--2\% centrality)~\cite{ATLAS:2019peb}.}
\label{fig:breathing}
\end{figure}

As we have demonstrated in Fig.~\ref{fig:vishnu}, the $v_2$-to-$v_3$ puzzle in ultracentral Pb+Pb collisions can be equally solved by introduce a static or breathing octupole deformation of \Pb\ .
In fact, it can be solved by any combination of the static octupole deformation $\avgoct$ and dynamic fluctuations $\sig{3}$ satisfying 
\begin{equation}  
\sqrt{\mean{\beta_3^2}}\equiv\sqrt{\avgoct^2 + \sig{3}^2} = 0.15.
	\label{eq:cons}
\end{equation}  
We now explore this by taking several combinations of $\avgoct$ and $\sig{3}$ that satisfy Eq.~(\ref{eq:cons}), using the initial-state quantities that are computationally efficient. In Fig.~\ref{fig:breathing}, we plot $\epsilon$ as a function of $\avgoct^2$ 
in ultracentral 0--1\% Pb+Pb collisions at $\snn=5.02$ TeV, calculated by the initial-state \trento\ model (open blue points, labeled by the right axis). The solid blue point depicting $\vvr$ from the full hydrodynamic \iebe\ simulations for the two limiting cases with $\sig{3}^{2}=0$ and 0.15 labeled by the extra right axis (the values of the two axes correspond to each other with the linear relationship $v_n \propto \epsilon_n$ for n=2,3 ). The experimental $\vvr$ data with error bars (gray band on the extra right axis) can be well reproduced. This implies that two-particle cumulant alone cannot distinguish the dynamic shape fluctuations from the static nuclear deformation of the colliding nuclei \Pb\ .

The degeneracy can be lifted by the four-particle cumulant. In Fig.~\ref{fig:breathing}, we also 
plot the initial-state ratio $\cce{3}$  calculated from \trento\ with various combinations of  $\avgoct$ and $\sig{3}$ (violet open points), and the final-state ratio $\cc{3}$ calculated from \iebe\  for the two limiting cases (the violet solid points). The four-particle cumulant ratios exhibit obvious sensitivity to $\sig{3}$.
Assuming a smooth correspondence from $\cce{3}$ to $\cc{3}$, the ATLAS data (arrow on the left vertical axis) prefers that the colliding nuclei \Pb\ is in a octupole breathing mode  with a root-mean-square amplitude $\sig{3}\approx0.06$ and the corresponding average octupole deformation is $\mean{\beta_3}\approx0.14$. 
This result is consistent with nuclear structure calculations from self-consistent configuration mixing  and multidimensional generator coordinate method~\cite{Yao:2016xkd,Henderson:2025scq}.
However, we emphasize that these values of octupole breathing should not be taken as precise extraction at current stage, but a qualitative imaging of  \Pb\ obtained from  relativistic heavy ion collisions. 
We also note that the $\epsilon_{3}$ to $v_{3}$ mapping breaks down in the presence of shape fluctuations. Further extraction of the shape fluctuation parameter requires a systematic investigation through Bayesian calibration and full hydrodynamic simulations.

{\em {\it Summary and outlook. }}
Relativistic heavy ion collisions take an instantaneous \textit{snapshot} of the overlap interaction zone of the two colliding nuclei and convert it by strong interactions in the created quark-gluon plasma into momentum anisotropies in the final state. The elliptical-to-triangular (\(v_2\)-to-\(v_3\)) anisotropic flow ratio in ultracentral \Pb+\Pb\ collisions at the Large Hadron Collider was found to be too small compared to hydrodynamic simulations with initial conditions containing the spherical shape of the colliding nuclei \Pb\ .  This puzzle itself can be solved by introducing a pure static octupole deformation (\(\avgoct = 0.15\) with \(\sigoct = 0\)), a pure fluctuating octupole (\(\avgoct = 0\) with \(\sigoct = 0.15\)), or anything in between. 
In this paper, we demonstrated that such ambiguity can be largely reduced by the four-particle cumulant $v_3\four$ measured in experiments.
Using the \iebe\ relativistic viscous hydrodynamic hybrid model simulations with the \trento\ initial condition, we show that solving the \(v_2\)-to-\(v_3\)  puzzle and simultaneously describing the $v_3\four$ data in ultracentral \Pb+\Pb\ collisions requires a dynamic octupole deformation with significant fluctuations--a breathing mode of $^{208}$Pb around an average octupole  shape. Experimental data seem to suggest values of \(\avgoct \approx 0.14\) and \(\sigoct \approx 0.06\).

In the present study, we implemented a simple Gaussian function for octupole shape fluctuations. 
However, shape fluctuations characterizing the time-dependent evolution of collective modes in nuclei originate, presumably, from nucleon-nucleon correlations~\cite{Butler:1996zz,Poves:2019byh}. Realistic nucleon-nucleon correlations exhibit significantly greater complexity than simple Gaussians could capture. Our work can be extended with more realistic profiles of nuclear structure and shape fluctuations. 
We have employed existing parameter sets in our initial-state \trento\ model and final-state \iebe\ hydrodynamic calculations and applied phenomenological parameterizations for nuclear shape and fluctuations in our present study to compare to experimental data. 
We have focused only on the $v_{2}/v_{3}$ ratio and ignored the individual $v_{2}$ and $v_{3}$ magnitudes as they can be tuned by hydrodynamic parameters. However, their tuned values will unlikely be proportional to each other. A full treatment would require Bayesian analysis incorporating nuclear density distributions~\cite{Giannini:2022bkn}, interplay between shape deformation and fluctuations (demonstrated in this work), hydrodynamic evolution and fluctuations~\cite{Kuroki:2023ebq}, QGP transport properties~\cite{Gardim:2014tya}, and technical issues like uncertainties in centrality definitions~\cite{Alqahtani:2024ejg} and estimations of non-collective flow effects~\cite{Feng:2024eos}.
Such an analysis remains a challenge to precisely extract the nuclear shape fluctuations and hydrodynamic parameters of the QGP from relativistic heavy ion collision data.
Future improved studies can be built upon our present work and will likely yield further insights into nuclear structure and fluctuations. 

The nuclear shape fluctuations investigated in this work  may only be discernible in relativistic heavy ion collisions whose yoctosecond timescale (\(\approx10^{-24}\)~s) preserves instant states of the nuclear wavefunction as snapshots~\cite{Dimri:2023wup,Giacalone:2024luz,STAR:2024wgy,Zhao:2024lpc,Giacalone:2024ixe,Mantysaari:2024uwn}.  
This allows the identification of the transient ``breathing mode"--time-dependent shape distortions--of the colliding nuclei and may enable novel probes of dynamic sub-nuclear structure such as nucleon clustering.
Such transient nuclear shapes may not be observable in low-energy nuclear reactions where these fluctuation effects are likely averaged out over the relatively long interaction times (\(\approx10^{-21}\)~s) by static equilibrium~\cite{Simenel:2020tdx}.
This underscores the emerging unique role of relativistic heavy ion collisions in probing novel physics in nuclear structure, complementary to and possibly beyond conventional low-energy nuclear reactions.

{\em {\it Acknowledgements. }} We thank Dr.~M. Luzum and Dr.~X. Wang for useful discussions. This work is supported in part by the National Natural Science Foundation of China under Grants No.~12275082 and No.~12035006 (H.X.), the National Science Foundation under grant number ACI-2004571 within the framework of the XSCAPE project of the JETSCAPE collaboration and by the U.S. Department of Energy, Office of Science, Office of Nuclear Physics, within the framework of the Saturated Glue (SURGE) Topical Theory Collaboration (W.Z.), the National Science Foundation under Grant No. 12247107 and No. 12075007 (X.X., S.Z. and H.S.), and the U.S. Department of Energy under Grant No.~DE-SC0012910 (F.W.).

{\em {\it Author contributions.}} H.X. conceived the idea, designed the study, made the plots, and wrote the initial manuscript; H.X. and W.Z. performed the Trento model simulations; H.X., D.X. and S.Z. performed the hydrodynamic calculations; F.W., H.S., and W.Z. provided critical feedback and revised the manuscript; All authors approved the final version.

{\em {\it Data availability.}} The data that support the findings of this article are not publicly available. The data are available from the authors upon reasonable request.

Notably concurrent with our arXiv submission, a recent study proposed to resolve the $v_2/v_3$ puzzle by considering the uniformity of nucleon and sub-nucleon structures in \Pb~\cite{Wang:2025cfu}.

\bibliography{ref2}

\end{document}